\documentclass[twocolumn,twocolappendix]{openjournal}

\usepackage[english]{babel}
\usepackage{amsmath,amssymb}
\usepackage{xcolor}
\usepackage{booktabs}
\usepackage{hyperref}

\definecolor{xlinkcolor}{cmyk}{1,1,0,0}
\hypersetup{
  unicode=true,
  pdfnewwindow=true,
  colorlinks=true,
  linkcolor=xlinkcolor,
  citecolor=xlinkcolor,
  filecolor=xlinkcolor,
  urlcolor=xlinkcolor
}

\begin{document}

\shorttitle{Long gamma-ray burst afterglows and the IGRB}
\shortauthors{S. Troitsky}

\title{Reassessing the contribution of high-energy afterglows of long gamma-ray bursts to the isotropic gamma-ray background after GRB~221009A}

\author{S.~Troitsky}
\email{st@inr.ac.ru}

\affiliation{Institute for Nuclear Research of the Russian Academy of Sciences,
  60th October Anniversary Prospect 7a,
  Moscow 117312,
  Russia}
\affiliation{Lomonosov Moscow State University,
  1-2 Leninskie Gory,
  Moscow 119991,
  Russia}

\begin{abstract}
Very-high-energy photons from gamma-ray bursts (GRBs) are absorbed by the extragalactic background light and partly reprocessed into the GeV band.  We update previous estimates of the long-GRB contribution to the Fermi-LAT isotropic gamma-ray background (IGRB), using a time-integrated high-energy afterglow template of GRB~221009A, anchored by GeV--TeV data, together with an explicitly specified scaling with prompt energy and beaming. Despite the observed multi-TeV afterglow, the cascade component reaches only \(1.6\times10^{-4}\) of the IGRB.  Including direct GeV emission raises the maximum fraction to \(1.1\times10^{-3}\), but makes the result more dependent on the population-averaged broadband template. Variations of the extragalactic background light, propagation implementation, intrinsic spectral cutoff, and population prescription do not change the conclusion: the multi-TeV detection of GRB~221009A does not make long GRBs a significant source of the IGRB.
\end{abstract}


\maketitle

\section{Introduction}
\label{sec:introduction}

The isotropic gamma-ray background (IGRB) measured by the Fermi Large Area
Telescope (LAT) contains the cumulative emission of unresolved extragalactic
sources together with any approximately isotropic residual foreground
component \citep{Ackermann2015,Fornasa2015}.  Electromagnetic cascades make the
IGRB particularly useful for testing source populations that inject photons
well above the energy range in which the Universe is transparent.  Pair
production, \(\gamma\gamma\to e^+e^-\), on the extragalactic background light
(EBL) and the cosmic microwave background (CMB), followed by inverse-Compton
scattering by the secondary electrons and positrons, transfers this energy to
lower-energy gamma rays \citep[for review and more references, see][]{BerezinskyKalashev2016}.

Gamma-ray bursts were recognized early as a possible contributor to this
cascade emission.  \citet{Casanova2007} explicitly calculated a prompt plus
scattered component, but the unknown TeV output had to be imposed as a model
assumption.  \citet{Murase2007} studied the closely related delayed pair echo
and its dependence on the cosmic infrared background and intergalactic
magnetic field.  From the GeV emission accessible to the Energetic Gamma Ray
Experiment Telescope (EGRET),
\citet{Ando2008} estimated a contribution of at least \(\sim10^{-4}\), and
probably \(\sim10^{-3}\), of the extragalactic background inferred from EGRET
data \citep{Sreekumar1998,Strong2004}.
These estimates became part of the standard inventory of diffuse gamma-ray
components \citep{Fornasa2015}, while the very-high-energy (VHE, above
100~GeV) normalization remained poorly
anchored observationally.  Later population calculations considered a
phenomenological high-energy prompt component \citep{Yao2020} and a physical
synchrotron plus synchrotron self-Compton (SSC) afterglow model
\citep{Min2024}; the
latter study found that GRB afterglows can contribute less than 10\% of the IGRB.

The situation changed with detections of VHE GRB afterglows and, in particular, with GRB~221009A.  The Water
Cherenkov Detector Array (WCDA) of the Large High Altitude Air Shower
Observatory (LHAASO) detected more than 64000 photons above 0.2~TeV during the
first 3000~s \citep{LHAASO2023}.  The independent LHAASO Kilometer Square
Array (KM2A) registered more than 140 photons above 3~TeV between 230 and
900~s, with reconstructed energies extending beyond 10~TeV
\citep{LHAASOKM2A2023}.  Carpet-3 subsequently reported a photon-like event
with an estimated primary energy of \(300^{+43}_{-38}\)~TeV at 4536~s
\citep{Carpet3_2025}.  The combined Fermi analysis provides a well-measured
high-energy history of the GRB \citep{Axelsson2025}, while these air-shower
data have been used to construct an intrinsic fluence spectrum extending
beyond the directly measured LHAASO range \citep{SatuninTroitsky2026}.
GRB~221009A is now used as a template in population forecasts for future
VHE detections \citep{Huang2026}.

Here we use these data to determine whether the observed VHE afterglow changes
the expected long-GRB contribution to the IGRB.  The intrinsic
\citet{GhirlandaSalvaterra2022} population is used as the baseline.  We report the
cascade separately from the direct-plus-cascade total because the former is
fixed primarily by the VHE energy budget, while the latter also depends on
assigning the GeV part of the GRB~221009A template to the population.  The total
is therefore an all-sky time-averaged broadband benchmark, not automatically
the fraction retained in an IGRB analysis after transient selection and
masking.  A
secondary calculation based on the older Population SYnthesis Code and
Hydrodynamic Emission model (PSYCHE) \citep{GhirlandaPop2013,GhirlandaPSYCHE2013,GhirlandaOrphan2015}
is retained to illustrate the population dependence.

The rest of the paper is organized as follows.  Section~\ref{sec:inputs}
defines the GRB~221009A spectral template and the baseline long-GRB
population.  Section~\ref{sec:propagation} describes photon propagation and
the observables, and Section~\ref{sec:results} presents the diffuse intensity.
Section~\ref{sec:history} compares the result with earlier estimates.
Sections~\ref{sec:discussion} and \ref{sec:conclusions} discuss the physical
interpretation and summarize the conclusions.  Population variations and
technical details are discussed in Appendices~\ref{app:populationchecks} and
\ref{app:technical}, respectively.

\section{Inputs and physical model}
\label{sec:inputs}

\subsection{GRB 221009A spectral template}
\label{sec:spectrum}

Our reference template
combines the Fermi high-energy afterglow shape with the multi-TeV behavior
measured by LHAASO and reconstructed in the high-energy fluence analysis.  The
seven smoothly broken power-law fits from the joint Fermi Gamma-ray Burst
Monitor (GBM) and LAT analysis cover
(280.6--435.6)~s and determine the early GeV--sub-TeV spectral shape
\citep{Axelsson2025}.  They are not treated as a measurement of the complete
event fluence.  Between 0.3 and 3~TeV, their duration-weighted spectral shape is
smoothly joined to the intrinsic high-energy fluence spectrum by
\citet{SatuninTroitsky2026}; this high-energy reconstruction is anchored
primarily by the WCDA spectrum below approximately 3~TeV.  The power-law
continuation above that range,
\begin{equation}
 E\,\frac{d{\cal F}}{dE}\propto E^{-0.315},
 \label{eq:tail}
\end{equation}
suggested  by
\citet{SatuninTroitsky2026}, is not a direct measurement up to the PeV
cutoff;  KM2A photons above 3~TeV and the Carpet-3 candidate event were used as consistency information for propagation scenarios.
Here and below, \(E\) is the observer-frame photon energy, \(z\) is the source
redshift, and \(E_s=(1+z)E\) is the source-frame energy.  We denote the intrinsic
(deabsorbed) energy fluence received at the Earth by \({\cal F}\).  The template
may be written as
\begin{equation}
 \frac{d{\cal F}}{dE}=A\,S(E),
 \label{eq:singleamplitude}
\end{equation}
where \(S(E)\) is its continuous GeV--PeV shape and \(A\) is a single
amplitude, fixed by
\begin{equation}
 \left.E\frac{d{\cal F}}{dE}\right|_{1\,\mathrm{TeV}}
 =4.236\times10^{-4}\ \mathrm{erg\,cm^{-2}}
 \label{eq:norm}
\end{equation}
for the intrinsic (0--2000)~s fluence.  A hard source-frame cutoff at 1~PeV is
used in the baseline calculation.  The integrated intrinsic fluence between
1~GeV and the cutoff is \(6.10\times10^{-3}\)~erg~cm\(^{-2}\), corresponding
to an isotropic-equivalent energy of approximately
\(3.52\times10^{53}\)~erg for the adopted luminosity distance.
For later population scaling, the source-frame isotropic-equivalent template
is obtained explicitly from the measured fluence at
\(z_*=0.15095\):
\begin{equation}
 \frac{dE_{\rm HE,iso}^{*}}{d\ln E_s}=
 \frac{4\pi d_L^2(z_*)}{1+z_*}
 \left.E\frac{d{\cal F}}{dE}\right|_{E=E_s/(1+z_*)}.
 \label{eq:sourceframe}
\end{equation}

The intrinsic spectrum of \citet{SatuninTroitsky2026} was introduced in a
study of axion-like-particle mixing and Lorentz-invariance violation, both of
which can reduce the effective pair-production opacity.  We instead propagate
the same source template with standard interactions.  At fixed injected
spectrum, any mechanism that allows a fraction of the photons to avoid pair
production transfers less energy to secondary pairs and therefore cannot
increase the cascade intensity calculated here.  It may increase the VHE direct
component, which is a different observable and is not modeled with new physics
in this work.

The three relevant time intervals have different observational meanings.
The (280.6--435.6)~s joint-fit interval fixes the early spectral shape; the
(0--2000)~s quantity in Eq.~(\ref{eq:norm}) is the reconstructed intrinsic VHE
fluence; and the much longer LAT result is obtained by integrating a fitted
light curve to 300~ks across separate visibility windows.  It is not a
continuous 300-ks exposure.  The Carpet-3 event at 4536~s
\citep{Carpet3_2025} lies outside the (0--2000)~s LHAASO fluence interval.  It is
used only as consistency information in the high-energy reconstruction and
does not supply an additional temporal normalization.  After applying the single VHE
anchor, the
template gives
\({\cal F}_{1-100\,{\rm GeV}}=3.524\times10^{-3}\)~erg~cm\(^{-2}\).
For orientation, the Fermi light-curve model gives
\((2.6\pm0.4)\times10^{-3}\)~erg~cm\(^{-2}\) in 0.1--100~GeV 
\citep{Axelsson2025}.  The template value is already 35\% higher although it
covers the narrower (1--100)~GeV interval, indicating a broadband-template
systematic of at least this order.  A direct component normalized
instead to the LAT fluence would be smaller; this ambiguity is not applied to
the cascade component and is smaller than the population dependence discussed
in Section~\ref{sec:discussion}.  The cascade-only contribution isolates the part
generated by VHE reprocessing and is less sensitive to the lower-energy
template uncertainty; we therefore report it separately.

The cutoff is a baseline regularization rather than an observationally
established feature; lowering the cutoff is among the cross checks discussed in Section~\ref{sec:results}.  We do not extrapolate the template beyond the PeV
scale: such an extension is not observationally motivated and would require
including the poorly constrained cosmic radio background in addition to the
EBL and CMB \citep{BerezinskyKalashev2016}.  Since the present calculation
already finds a sub-percent IGRB
contribution, this high-energy continuation is not a leading uncertainty of
the diffuse-background result.

\begin{figure}[t]
\centering
\includegraphics[width=\columnwidth]{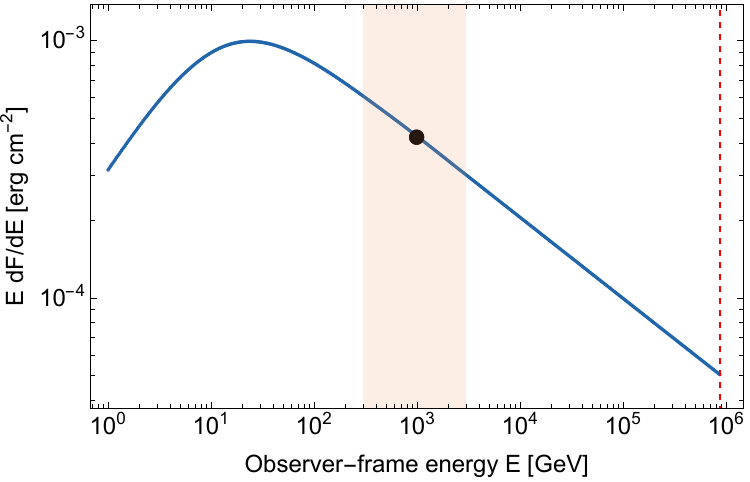}
\caption{Reference GRB~221009A intrinsic fluence template.  The GeV--sub-TeV shape is
based on Fermi spectral fits \citep{Axelsson2025};
the single normalization (black dot) and the VHE tail use the reconstructed intrinsic
0--2000~s fluence, anchored primarily by WCDA below approximately 3~TeV, with
KM2A and Carpet-3 providing high-energy consistency information
\citep{LHAASO2023,LHAASOKM2A2023,Carpet3_2025,SatuninTroitsky2026}.
The two spectral segments are smoothly joined in the pink band.  The dashed line illustrates the cutoff of
1~PeV in the source galaxy frame (0.869~PeV in the observer frame for GRB~221009A).}
\label{fig:source}
\end{figure}

The resulting input fluence spectrum template is presented in Fig.~\ref{fig:source}.

\subsection{Empirical check of the time-integrated VHE template}
\label{sec:vhetemplate}

The cascade calculation requires the time-integrated emitted spectrum rather
than a model of the VHE light curve.  We therefore compare published intrinsic
VHE fluence spectra without extrapolating unobserved time intervals.  For event
\(i\) we define
\begin{equation}
 \Psi_i(E_s)=\frac{1}{E_{\gamma,{\rm iso},i}}
 \frac{dE_{{\rm VHE,iso},i}}{d\ln E_s},
 \label{eq:psi}
\end{equation}
where \(E_{\gamma,{\rm iso},i}\) is the prompt isotropic-equivalent energy and
\(dE_{{\rm VHE,iso},i}/d\ln E_s\) is the isotropic-equivalent VHE energy per
logarithmic source-frame energy interval.  The common source-frame interval
0.5--1~TeV is directly covered by the Major Atmospheric Gamma Imaging
Cherenkov (MAGIC) spectrum of GRB~190114C \citep{MAGIC190114C}, the High
Energy Stereoscopic System (H.E.S.S.) spectrum of GRB~190829A
\citep{HESS190829A}, and our template of GRB~221009A.

Direct integration of the published MAGIC bins gives
$E_{\rm VHE,iso}(0.3\text{--}1\,\mathrm{TeV})=3.92\times10^{51}$~erg for
GRB~190114C, reproducing the published estimate of order
$4\times10^{51}$~erg.  Integration of energy fluxes presented in Table~S2 of \citet{HESS190829A} over the three actually observed intervals gives $8.88\times10^{48}$~erg in (0.2--4)~TeV for GRB~190829A.  No temporal extrapolation is made; hence these observed-window energies are lower bounds on the full-event VHE output.

\begin{table*}[t]
\caption{Time-integrated VHE spectral-energy normalization in the common
source-frame (0.5--1)~TeV band.  The ratios for GRB~190114C and GRB~190829A use
only their actually observed VHE intervals and therefore should not be
interpreted as population-mean correction factors.}
\label{tab:vhetemplate}
\centering
\begin{tabular}{@{}lccc@{}}
\toprule
Event & Instrument & $E_{\rm VHE,iso}(0.5{-}1\,\mathrm{TeV})/E_{\gamma,\rm iso}$ & Ratio to GRB~221009A \\
\midrule
GRB~221009A & LHAASO, Fermi LAT & $1.946\times10^{-3}$ & 1.00 \\
GRB~190114C & MAGIC & $9.212\times10^{-3}$ & 4.73 \\
GRB~190829A & H.E.S.S. & $1.048\times10^{-2}$ & 5.38 \\
\bottomrule
\end{tabular}
\end{table*}

Writing the intrinsic photon spectrum locally as
\(dN_\gamma/dE_s\propto E_s^{-\Gamma}\), the central photon indices in the
common TeV region are also
similar: $\Gamma=2.22$ for GRB~190114C, $\Gamma=2.07$ for GRB~190829A,
and a local effective $\Gamma\simeq2.30$ for GRB~221009A.  Thus the currently useful
cross-event difference is primarily one of normalization rather than a
qualitatively different TeV spectral slope.  Other VHE GRBs
do not provide comparable leverage above a few TeV, so the population
diversity of the multi-TeV tail is not measured by present data.

The two comparison bursts lie by factors 4.73 and 5.38 above, rather than
below, the GRB~221009A scaling in the common band (Table~\ref{tab:vhetemplate}).
Thus GRB~221009A is not uniquely VHE-efficient within the detected sample.
This is a consistency check, not a measurement of the population mean: the
VHE sample is strongly biased toward bright events and favorable observing
conditions.  The second H.E.S.S. GRB catalogue independently finds that
VHE-detected bursts tend to have luminous X-ray afterglows, favorable
redshifts, and favorable observing conditions \citep{HESSCatalogue2026}.
Published nondetections over finite follow-up windows likewise do not provide
model-independent upper limits on the full time-integrated VHE fluence unless
an additional temporal model is imposed.  Unlike triggered MAGIC and H.E.S.S.
follow-up, LHAASO operates continuously, so the observation of GRB~221009A was
obtained from regularly taken wide-field data.  The event is nevertheless
exceptional in fluence and proximity, and no population distribution is
inferred from these three detections. 

\subsection{Long-GRB population}
\label{sec:population}
For the baseline population model, we use the long-GRB cosmic history of
\citet{GhirlandaSalvaterra2022}.  We denote its comoving event-rate density by
\(\rho(z)\); the local true rate is
\(\rho_0=79\ \mathrm{Gpc^{-3}\,yr^{-1}}\), with redshift dependence
\begin{equation}
 \rho(z)=\rho_0\frac{(1+z)^{3.33}}{
 1+[(1+z)/3.42]^{6.21}} .
 \label{eq:rate}
\end{equation}
The luminosity function and its moderate redshift evolution are implemented
together with the empirical relations between isotropic luminosity, peak
energy, isotropic energy, and the two-sided beaming fraction.  We refer to
this model as GS.

The mapping from GRB~221009A to every synthetic burst is a source assumption,
not an output of the population model.  If \(E_{\gamma,{\rm iso}}^*\) is the
prompt isotropic-equivalent energy of GRB~221009A and
\(E_{\gamma,{\rm iso}}\) that of a population member, we assign
\begin{equation}
 \frac{dE_{\rm HE,true}}{d\ln E_s}=
 f_{b,{\rm VHE}}
 \frac{E_{\gamma,{\rm iso}}}{E_{\gamma,{\rm iso}}^*}
 \frac{dE_{\rm HE,iso}^{*}}{d\ln E_s}.
 \label{eq:population_scaling}
\end{equation}
For the numerical scaling we use
\(E_{\gamma,{\rm iso}}^*=1.0\times10^{55}\)~erg for GRB~221009A,
following \citet{OConnor2023}.  
Equation~(\ref{eq:population_scaling}) assumes a universal high-energy to
prompt-energy ratio and a universal source-frame spectral shape, including a
fixed source-frame cutoff.  Population scatter around either relation is not
constrained by present VHE data.

For the beaming bookkeeping, let \(\theta_j\) be the jet half-opening angle and
\(f_b=1-\cos\theta_j\) the two-sided top-hat beaming fraction.  If
\(R_{\rm true}\) is the rate of all
jets and \(R_{\rm on}=f_bR_{\rm true}\) the rate whose prompt beams point
towards us, while \(E_{\rm true}=f_bE_{\rm iso}\), then
\begin{equation}
 R_{\rm on}E_{\rm iso}=R_{\rm true}E_{\rm true}.
 \label{eq:beaming_cancel}
\end{equation}
Thus the same mean emissivity is obtained from an orientation-selected rate
with isotropic-equivalent energy or from the intrinsic rate with the true
collimation-corrected energy, provided the conventions are not mixed.  More
generally, for randomly oriented jet axes, convolution with any normalized
angular redistribution kernel for the secondary cascade preserves the sky
monopole (the zeroth angular moment) of the ensemble-averaged intensity.  Intergalactic magnetic
fields therefore affect pair-echo delays, angular profiles, source
associations, and masking, but not the long-term all-sky mean simply by
redistributing directions \citep{Murase2007,BerezinskyKalashev2016}.

The GS construction is consistent with Eq.~(\ref{eq:beaming_cancel}):
\citet{GhirlandaSalvaterra2022} quote an intrinsic rate accounting for
collimation, and we weight each event by \(f_bE_{\rm iso}\).  A remaining
source-physics assumption is that the VHE afterglow is beamed over
approximately the same solid angle as that used for the prompt/early
afterglow correction.  This relation is not measured for the population.
Constraints from the observed ratio of early X-ray afterglow to prompt energy
nevertheless disfavor arbitrarily different angular structures for the two
components \citep{BeniaminiNakar2019}.  For GRB~221009A itself the
\(\sim0.8^\circ\) VHE jet scale inferred from the LHAASO light curve
\citep{LHAASO2023} is close to the \(\theta_c\simeq0.021\) rad
(\(\simeq1.2^\circ\)) afterglow core inferred by \citet{GillGranot2023}.
Structured-jet fits allow much wider kinetic-energy wings
\citep{OConnor2023,Sato2025}, but the existence of kinetic energy at large
angles does not imply the same TeV radiative efficiency there.  We therefore
use \(f_{b,\rm VHE}=f_{b,\rm prompt}\) as the baseline.  If the VHE-emitting
core is systematically narrower while its on-axis isotropic-equivalent
spectrum is unchanged, the diffuse result scales down approximately as
\(f_{b,\rm VHE}/f_{b,\rm prompt}\).

While we use GS as the baseline population model, the sensitivity to other
population prescriptions is discussed separately in
Appendix~\ref{app:populationchecks}.

\section{Propagation and observables}
\label{sec:propagation}
We calculate the ensemble-averaged intensity produced by direct transmission
and electromagnetic reprocessing.  The evolving EBL is taken from the
observational model of \citet{SaldanaLopez2021}, with the CMB added explicitly.
Its central realization defines the baseline; the published low- and
high-density realizations define the GS-only EBL envelope.  Model A of
\citet{Finke2022} provides an independent EBL comparison.  Cascade propagation
is calculated with CRbeam \citep{Kalashev2023}.  We denote the observed
differential photon intensity by \(I(E)\).  For a comoving differential photon
emissivity,
\begin{equation}
 Q(E_s,z)=\rho(z)\left\langle\frac{dN_\gamma}{dE_s}\right\rangle,
 \label{eq:emissivity}
\end{equation}
measured in photons per comoving volume, source-frame time, and source energy,
the number-intensity weight of a source-energy and redshift cell is
\begin{equation}
 d{\cal J}_{\rm inj}=\frac{c\,Q(E_s,z)\,dE_s\,dz}{4\pi H(z)(1+z)}.
 \label{eq:intensity}
\end{equation}
The propagation response redistributes this injected-cell weight over observed
energy to produce \(I(E)\).  Here \(c\) is the speed of light and \(H(z)\) is
the Hubble expansion rate.
We adopt \(H_0=67.4\)~km~s\(^{-1}\)~Mpc\(^{-1}\), with present matter and
dark-energy density parameters \(\Omega_m=0.315\) and
\(\Omega_\Lambda=0.685\), respectively \citep{Planck2020}.

The baseline population integral is truncated at \(z_{\max}=6\), the upper
boundary of the tabulated Salda\~na--Lopez EBL model.  No EBL extrapolation
beyond \(z=6\) is introduced.  To estimate the omitted high-redshift
contribution, we also extend the source population to \(z_{\max}=10\); in this
sensitivity calculation the tabulated EBL contribution is omitted above
\(z=6\), while the CMB continues to evolve.  

The reference result uses the central EBL realization and neglects
intergalactic magnetic-field energy losses.  Thus the statement that angular
redistribution preserves the mean intensity applies when magnetic deflection
only redistributes arrival directions and times, inverse-Compton losses remain
dominant, and the signal is averaged over a sufficiently long interval.  Very
long delays and synchrotron cooling in stronger fields are outside the scope
of the present mean-intensity calculation
\citep{Murase2007,BerezinskyKalashev2016}.

The direct, unscattered component is computed deterministically from the optical depth.  The cascade component contains only photons that have undergone at least one interaction, which avoids double counting the direct flux.  We therefore define
\begin{equation}
 I_{\rm total}(E)=I_{\rm direct}(E)+I_{\rm cas}(E),
 \label{eq:components}
\end{equation}
and compare both \(I_{\rm cas}\) and \(I_{\rm total}\) with the IGRB.  The cascade-only comparison is less sensitive to how bright transient intervals are represented in a diffuse-data product, whereas the total is the physical all-sky time average under the adopted broadband source model.  Its ratio to the measured IGRB is therefore a benchmark on the same intensity scale, not a literal measurement of the unresolved GRB fraction in that data product.

The 50-month Fermi-LAT IGRB analysis \citep{Ackermann2015} does not document a
global time veto for GRBs.  Short intervals around four bright bursts were
excluded when constructing the Second Fermi-LAT Source Catalog (2FGL)
\citep{Nolan2012}, but that choice does not uniquely specify how a
population-averaged transient contribution maps onto the later IGRB product.
This observational ambiguity is another reason to report the cascade-only
flux separately from the direct-plus-cascade total.

The propagation calculation was validated with monoenergetic injections,
changes of the interpolation grid, and an independent point-kernel calculation
with $\gamma$-Cascade \citep{CapanemaBlanco2025}.  With identical physical
conventions in the matched \(z_{\max}=10\) calculation, the latter gives
a cascade peak of \(1.8\times10^{-4}\) of the
IGRB and a cascade-only intensity 26.78\% higher when integrated over
0.1--820~GeV.  Its total intensity is about 6\% higher over the same range and
differs by less than 3\% at 1--10~TeV; neither result was normalized to the
other.  We retain the cascade-only difference as an inter-code propagation
systematic rather than hiding it inside the smaller total-component
difference.  The transport grids, statistical accuracy, and component-level
comparisons are described in Appendix~\ref{app:technical}.

\section{Results}
\label{sec:results}
We fold the predicted spectra into the published Fermi-LAT energy bins and
define the bin ratio
\begin{equation}
 R_{X,k}=\frac{\int_{\Delta E_k}I_X(E)\,dE}
 {\Phi_{{\rm IGRB},k}},
 \qquad X\in\{{\rm cas,total}\},
 \label{eq:binratio}
\end{equation}
where \(\Phi_{{\rm IGRB},k}\) is the measured bin-integrated IGRB intensity.
Both GS ratios peak in the 18--26~GeV bin \citep{Ackermann2015}.  The more
robust cascade-only result is
\begin{equation}
 R_{\rm cas,peak}^{\rm GS}\approx 1.6\times10^{-4},
 \label{eq:gscascade}
\end{equation}
whereas the direct-plus-cascade broadband benchmark gives
\begin{equation}
 R_{\rm total,peak}^{\rm GS}\approx 1.1\times10^{-3}.
 \label{eq:gsmax}
\end{equation}
About 85\% of the total in the peak bin is direct emission.
We remind that Eq.~(\ref{eq:gscascade}) tests VHE
energy reprocessing, while Eq.~(\ref{eq:gsmax}) additionally depends on the
low-energy shape of the same continuous broadband template.  The quoted
ratios characterize the astrophysical scale rather than numerical precision.
As we discuss below, uncertainties in the GRB rate and energy distribution are much
larger than the Monte Carlo sampling error, but cannot be represented by
rescaling the local rate alone because the population parameters are
correlated.

To assess the sensitivity of our conclusions to the assumed PeV cutoff, we performed a test with the cutoff value just above the highest energy of LHAASO-detected photons from GRB~221009A, that is $E=20$~TeV, corresponding to $E_s=23$~TeV. The low-cutoff test gives
\(R_{\rm cas,peak}=1.15\times10^{-4}\), 70\% of the 1-PeV baseline.  The corresponding total peak is
\(1.02\times10^{-3}\), 95\% of baseline, because it is dominated by direct GeV
photons.  Thus even a cutoff close to the directly constrained multi-TeV range
changes the cascade estimate by only 30\% and does not weaken the
conclusion.  In the matched \(z_{\max}=10\) validation calculation with 
$\gamma$-Cascade, we obtain \(R_{\rm cas,peak}=1.78\times10^{-4}\); the propagation implementations therefore span cascade peaks of approximately
\((1.6\text{--}1.8)\times10^{-4}\).

Extending the redshift integral from the baseline \(z_{\max}=6\) to
\(z_{\max}=10\) raises the cascade and total peak ratios by 0.47\% and 0.45\%,
respectively; their (0.1--820)~GeV integrated intensities increase by less than
1\%.  In this estimate, the \(z>6\) contribution is below the
percent level; its uncertain EBL treatment therefore cannot affect the
conclusion.

\begin{figure}[t]
\centering
\includegraphics[width=\columnwidth]{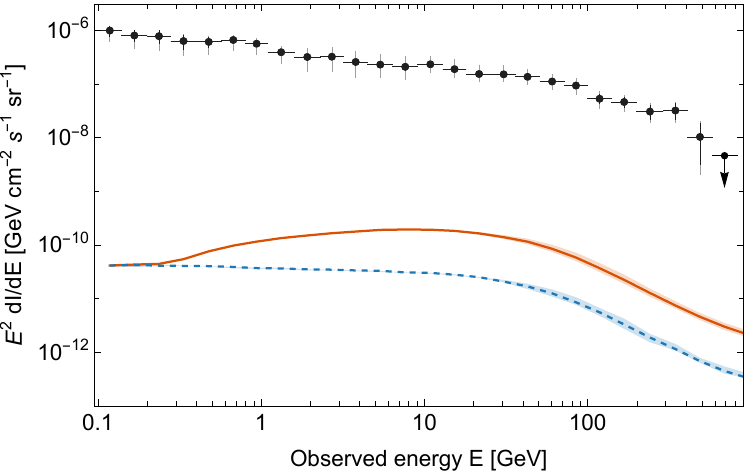}
\caption{Long-GRB contribution to IGRB compared with the Fermi-LAT model-A (black) measurements \citep{Ackermann2015}.  The GS cascade-only result shown in blue (dashed) is less model dependent than the direct-plus-cascade total (orange).  Each shaded band is the envelope for the Salda\~na--Lopez central, low, and high EBL densities.  Black inner error bars contain the model A published non-foreground uncertainties; gray outer bars also include the Galactic foreground range.}
\label{fig:result}
\end{figure}

We compare the diffuse GRB spectrum with the Fermi-LAT model-A IGRB in
Fig.~\ref{fig:result}.  The outer error bars show the published Galactic
foreground range, including alternative foreground models and high-latitude
normalization tests, rather than treating models A--C as three independent
measurements \citep{Ackermann2015}.  This foreground systematic and the
population uncertainty both exceed the sampling uncertainties documented in
Appendix~\ref{app:crbeam}.

Changing the EBL model leaves the conclusion unchanged.  The three
Salda\~na--Lopez realizations give about 0.016\% of the IGRB for the cascade
peak and (0.10--0.11)\% for the total.  Figure~\ref{fig:result} shows the
envelope of these realizations.  Replacing the
Salda\~na--Lopez model by the Finke model gives peak ratios of 0.017\%
and 0.11\%, respectively.

\begin{figure}[t]
\centering
\includegraphics[width=\columnwidth,
  trim=20 0 5 0,
  clip]{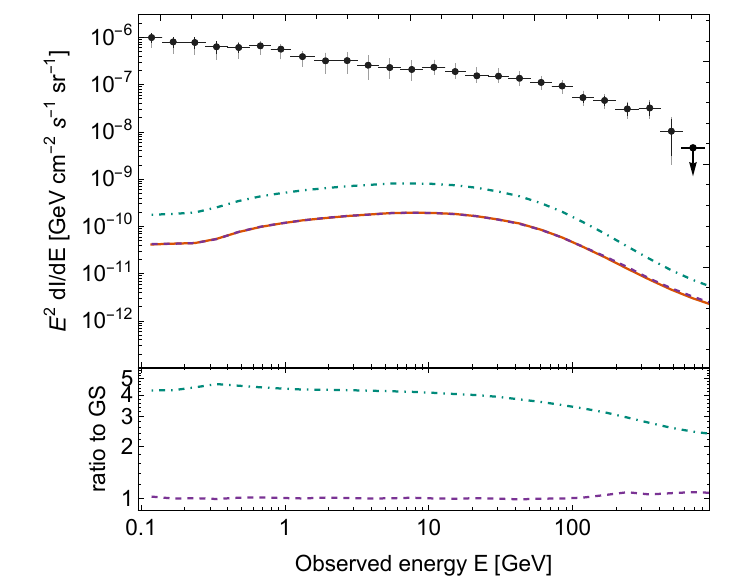}
\caption{Model variations of the direct-plus-cascade total.  The GS plus
Salda\~na--Lopez curve (orange, solid) is the baseline.  The Finke curve
(violet, dashed) changes only the EBL, whereas PSYCHE (green, dash-dotted)
changes the population history and the normalization assigned to a typical
burst.  The lower panel shows ratios to the GS baseline.  Fermi-LAT symbols
are as in Fig.~\ref{fig:result}.}
\label{fig:variations}
\end{figure}

Figure~\ref{fig:variations} separates the population and EBL dependences.  The
central PSYCHE synthesis raises the total by a factor of about four for the
reasons discussed in Section~\ref{sec:psyche} and detailed in
Appendix~\ref{app:psyche}.  Changing the EBL model from Salda\~na--Lopez to
Finke at fixed GS population has a much smaller effect.
The spectrum decreases monotonically in the TeV range.  The smallness of the IGRB contribution follows directly from Eqs.~(\ref{eq:gscascade}) and (\ref{eq:gsmax}).

\section{Comparison with earlier estimates}
\label{sec:history}

Table~\ref{tab:history} summarizes the conceptual difference between the
present calculation and representative earlier studies.  The crucial change
is not the cascade mechanism itself, which was already discussed explicitly
by \citet{Casanova2007} and \citet{Murase2007}, but the empirical information
now available for the intrinsic VHE normalization.
\begin{table*}[t]
\caption{Representative estimates of the GRB contribution to diffuse GeV--TeV emission.}
\label{tab:history}
\centering
\begin{tabular}{@{}lll@{}}
\toprule
\parbox[t]{0.14\textwidth}{Study} &
\parbox[t]{0.22\textwidth}{Component} &
\parbox[t]{0.25\textwidth}{VHE normalization} 
\\
\midrule
\parbox[t]{0.14\textwidth}{\citet{Casanova2007}} &
\parbox[t]{0.22\textwidth}{Prompt plus reprocessed TeV emission} &
\parbox[t]{0.25\textwidth}{Model assumption; an optimistic TeV energy output was adopted} 
\\[26pt]
\parbox[t]{0.14\textwidth}{\citet{Murase2007}} &
\parbox[t]{0.22\textwidth}{Delayed pair echo / regenerated emission} &
\parbox[t]{0.25\textwidth}{Intrinsic TeV spectra specified by source models} 
\\[16pt]
\parbox[t]{0.14\textwidth}{\citet{Ando2008}} &
\parbox[t]{0.22\textwidth}{Prompt and afterglow GeV emission} &
\parbox[t]{0.25\textwidth}{Constrained with EGRET data and SSC modeling} 
\\[16pt]
\parbox[t]{0.14\textwidth}{\citet{Yao2020}} &
\parbox[t]{0.22\textwidth}{Phenomenological high-energy prompt emission} &
\parbox[t]{0.25\textwidth}{Population correlations and an extrapolated VHE component}
\\[16pt]
\parbox[t]{0.14\textwidth}{\citet{Min2024}} &
\parbox[t]{0.22\textwidth}{Synchrotron and SSC afterglow emission} &
\parbox[t]{0.25\textwidth}{Swift X-ray and Fermi-LAT afterglow energetics}
\\[16pt]
\parbox[t]{0.14\textwidth}{This work} &
\parbox[t]{0.22\textwidth}{Afterglow from GeV to VHE} &
\parbox[t]{0.25\textwidth}{GRB~221009A broadband VHE afterglow template} 
\\
\bottomrule
\end{tabular}
\end{table*}

\begin{figure}[t]
\centering
\includegraphics[width=\columnwidth]{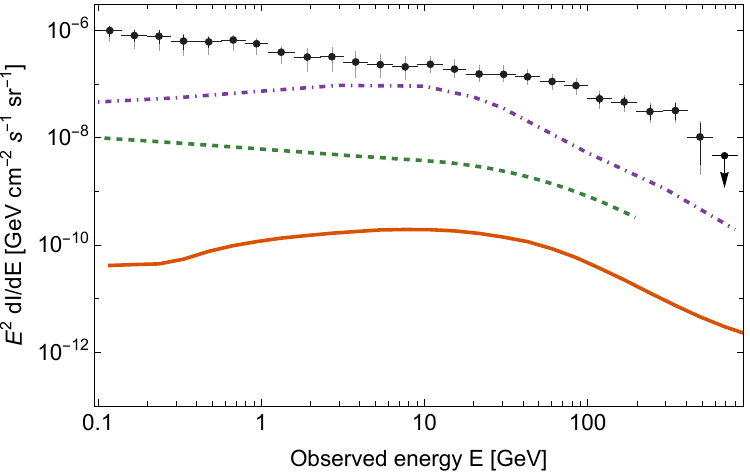}
\caption{Comparison with earlier spectral calculations.  Shown are the GS
total from this work (orange), the Casanova et al.\ prompt plus scattered
sum (violet dot-dashed), the Murase et al.\ model-A primary-plus-delayed
total (green dashed), and Fermi-LAT IGRB model A as in
Fig.~\ref{fig:result}.  The integrated result of Ando et al., for which no
corresponding spectrum was published, is compared in the text.}
\label{fig:historical}
\end{figure}

For the spectral comparison in Fig.~\ref{fig:historical}, we digitized the
published prompt-plus-scattered ``sum'' curve from Fig.~5 of
\citet{Casanova2007} and the model-A primary and delayed curves from Fig.~14
of \citet{Murase2007}; the latter two are added arithmetically.  Around
1--30~GeV, the deliberately optimistic Casanova curve exceeds the GS baseline
by roughly 2.4--2.8 orders of magnitude.  Murase model A is shown as a
representative intrinsic-spectrum choice.

The more recent calculation by \citet{Min2024}, specifically concerned with
afterglows, varied the microphysical parameters of a synchrotron plus SSC
model normalized with Swift X-ray and Fermi-LAT observations and found a
contribution below 10\% of the IGRB.  This broad upper bound is consistent with
our result, but it addresses a different question.  Here the high-energy
normalization is fixed directly by the GRB~221009A fluence template and the
transmitted and cascade components are propagated separately.  The
phenomenological prompt-emission calculation of \citet{Yao2020} is likewise
not used to normalize our afterglow template.

\citet{Ando2008} did not publish a spectrum suitable for the plot, but their
Table~2 gives an intensity integrated over 30~MeV--30~GeV.  Summing the
nominal prompt and afterglow terms gives
\((1.1\text{--}2.1)\times10^{-9}\)~GeV~cm\(^{-2}\)~s\(^{-1}\)~sr\(^{-1}\) for their
models A--C; the largest prompt inverse-Compton contribution considered there
raises this range by a factor of $\sim 5$.  Our GS baseline,
integrated over its supported (0.1--30)~GeV range, is \(7.3\times10^{-10}\)~GeV~cm\(^{-2}\)~s\(^{-1}\)~sr\(^{-1}\), or 0.028\% of the Fermi model-A IGRB in that range.  Thus the
nominal Ando intensities exceed our result by factors of approximately (1.5--3), whereas their
inverse-Compton-enhanced values are larger by factors about (7--13).  Their quoted
fractions, at least 0.01\% and possibly about 0.1\%, referred to the EGRET
extragalactic background \citep{Sreekumar1998,Strong2004}; the percentages
should not be compared directly with Fermi-LAT values because both the
background definition and the lower energy boundary differ.

\section{Discussion}
\label{sec:discussion}

\subsection{Baseline interpretation}
\label{sec:baseline}

Our principal result is the small GRB cascade component.  Replacing the
formerly free TeV energy budget by the GRB~221009A template gives a peak ratio
of only \(1.6\times10^{-4}\) to the IGRB; the independent transport result is
\(1.8\times10^{-4}\).  This statement depends on the VHE
emissivity and propagation, but not on whether direct photons from catalogued
transients are retained in a particular diffuse-data product.  The
direct-plus-cascade peak ratio, \(1.1\times10^{-3}\), answers the broader but
more model-dependent question of the full all-sky time-averaged broadband
output.  It is dominated by direct photons near 20~GeV and should not be read
as a literal unresolved-source fraction in the Fermi-LAT IGRB product.

The published Salda\~na--Lopez EBL-density variants keep the cascade peak near
0.016\% and the total peak within (0.10--0.11)\%.
The independent Finke EBL calculation is similarly close to the central
result.  Thus the EBL choice does not compete with the source-rate and
energy-calibration uncertainties for this population estimate.

The one-amplitude construction in Eq.~(\ref{eq:singleamplitude}) and the
population mapping in Eq.~(\ref{eq:population_scaling}) are important for
interpreting this result.  The early (280.6--435.6)~s spectral fits, the intrinsic
(0--2000)~s VHE fluence, and the LAT light-curve integral to 300~ks are different
temporal--spectral reconstructions.  Their GeV mismatch is at least 35\% and is
retained as a broadband-template systematic rather than used to renormalize the
GeV and VHE branches independently.  The appropriate next source model is a
population distribution of continuous broadband spectra, constrained jointly
at low and high energy.  The LAT GRB catalogue already demonstrates large
dispersion and strong high-energy selection effects \citep{Ajello2019}, whereas
the VHE sample is still small and Malmquist-biased.

The observed-window comparison in Table~\ref{tab:vhetemplate} also supplies a
simple stress test.  Multiplying the entire GRB~221009A high-energy template by
the largest tabulated factor, 5.38, would raise the baseline cascade peak only
to 0.088\% of the IGRB.  A factor of $\approx 61$ in the
population-averaged VHE normalization would be required for the cascade alone
to reach 1\%.  Both rescalings are deliberately more favorable to a cascade
than the lower-cutoff or steeper-tail alternatives, which reduce it.

\subsection{Origin of the PSYCHE--GS difference}
\label{sec:psyche}
The PSYCHE calculation (Appendix~\ref{app:psyche}) uses the same spectral
shape and the same propagation responses as the baseline, so the
factor-of-four difference cannot originate from cascade propagation.  It is
set by the population emissivity.  In the central PSYCHE calibration, the mean
energy assigned to the GeV--PeV template is \(8.03\times10^{49}\)~erg per
burst and the local rate is 6.35~Gpc\(^{-3}\)~yr\(^{-1}\), giving a local
high-energy production density of
\(5.10\times10^{50}\)~erg~Gpc\(^{-3}\)~yr\(^{-1}\).  The corresponding GS
population average gives \(3.10\times10^{50}\)~erg~Gpc\(^{-3}\)~yr\(^{-1}\)
at \(z=0\), so PSYCHE is already higher by a factor of 1.64 locally.  Its
redshift evolution gives greater weight to \(z\sim1\text{--}5\), supplying a further
factor of about 2.5 after cosmological redshifting.  The resulting ratio,
about 4.1, agrees with the ratio of the propagated intensities.

The larger PSYCHE energy per event follows from its calibration rather than
from its local rate.  Our baseline calculation scales the
GRB~221009A template, for which the calculation uses the fiducial prompt
isotropic-equivalent energy of \(1.0\times10^{55}\)~erg
\citep{OConnor2023}, to the distribution of ordinary GRB
prompt energies and beaming fractions.  PSYCHE instead applies a
high-energy-template efficiency to the mean kinetic
energy of its synthetic bursts.  These are different source-population
assumptions.  The PSYCHE result therefore provides a sensitivity test but is
neither an independent propagation calculation nor an uncertainty estimate
for GS.  Its central peak ratios are 0.066\% for the cascade and 0.42\% for the
total.  Combining the two deliberately high-energy calibrations in
Table~\ref{tab:psyche} gives the most extreme tested PSYCHE case, 0.18\% for
the cascade and 1.14\% for the broadband total.  This extreme is a diagnostic
scenario, not an uncertainty band around GS.

\subsection{Implications}
\label{sec:implications}
The small standard-afterglow contribution leaves room to constrain a separate
hidden TeV--PeV or hadronic component with the residual IGRB after known source
classes have been included.  For photohadronic models this constraint 
complements IceCube limits on GRB neutrinos \citep{IceCubeGRB2022}, although a
joint gamma--neutrino analysis would require explicit assumptions about source
opacity, baryon loading, and escape fractions.  Present uncertainties in the
GRB high-energy luminosity function and its redshift evolution preclude a
robust joint population limit.

An angular-power prediction would require an explicit transient time window,
source mask, luminosity-function tail, and magnetic-delay model, and is beyond
the monopole calculation performed here.  More direct tests of standard
afterglows are searches for delayed or extended cascade emission from
individual bright GRBs and improved measurements of the VHE-efficiency
distribution with the Cherenkov Telescope Array Observatory (CTAO)
\citep{InoueCTA2013,HESSCatalogue2026}.

\section{Conclusions}
\label{sec:conclusions}
The multi-TeV afterglow of GRB~221009A does not promote long GRBs to a
significant source of the Fermi IGRB.  For the GS population, the cascade peak is 0.016\% of the background; an independent propagation implementation gives 0.018\%, despite a 27\% difference in the (0.1--820)~GeV cascade-only
integral.  The all-sky time-averaged direct-plus-cascade contribution peaks at
0.11\% of the IGRB, but is not directly an unresolved-source fraction.  A
lower source cutoff, a steeper VHE tail, or a narrower VHE beam only lowers
the cascade.  The central PSYCHE diagnostic gives 0.066\% cascade and 0.42\%
total, while the most extreme tested calibration gives 0.18\% and 1.14\%. EBL and propagation variations
do not change the conclusion that standard long-GRB high-energy
afterglows do not constitute a significant component of the IGRB.

\begin{appendix}
\section{Variations in the population model}
\label{app:populationchecks}
\subsection{PSYCHE all-orientation synthesis}
\label{app:psyche}
The secondary calculation uses the PSYCHE synthesis introduced in
Section~\ref{sec:introduction} and its explicit off-axis extension
\citep{GhirlandaOrphan2015}.  Each synthetic event has a jet opening angle,
initial Lorentz factor, redshift, and random viewing angle.  About 2.4\% of the
parent population points toward the observer.  The adopted comoving event-rate
density is
\begin{equation}
 \rho_{\rm P}(z)\propto
 \frac{0.0157+0.118z}{1+(z/3.23)^{4.66}}(1+z)^{1.7},
 \qquad z\leq10,
 \label{eq:psyche_rate}
\end{equation}
normalized with the PSYCHE orphan/on-axis population.  In the cosmology used
here, this corresponds to the local rate
\(\rho_{0,\rm P}=6.35\ {\rm Gpc^{-3}\,yr^{-1}}\).
The rate model remains defined to \(z=10\) for this normalization, whereas the
quoted diffuse PSYCHE spectra use the same \(z_{\max}=6\) integration limit as
the GS baseline.  Extending the PSYCHE fold to \(z=10\) raises its peak ratios
by less than 1\% and its 0.1--820~GeV integrated cascade and total intensities
by 1.2\% and 1.6\%, respectively.

The afterglow calibration does not use the original fixed prompt efficiency.
In PSYCHE, \(E'_\gamma=1.5\times10^{48}\)~erg is the prompt radiated energy
in the comoving frame and \(\Gamma_0\) is the initial bulk Lorentz factor.
Thus \(E_{\gamma,\rm true}=\Gamma_0E'_\gamma\), and
\(\langle\Gamma_0\rangle=274\) gives
\(\langle E_{\gamma,\rm true}\rangle=4.11\times10^{50}\)~erg.  Here
\(E_{\gamma,\rm true}\) and \(E_{K,\rm true}\) are the collimation-corrected
prompt radiated and blast-wave kinetic energies.  The prompt radiative
efficiency is
\(\eta_\gamma=E_{\gamma,\rm true}/(E_{\gamma,\rm true}+E_{K,\rm true})\).
For the mean long-GRB value \(\eta_\gamma=0.26\) inferred by
\citet{Aksulu2022}, the central kinetic energy is
\(\langle E_{K,\rm true}\rangle=1.17\times10^{51}\)~erg.  The high-energy
scenario in Table~\ref{tab:psyche} uses \(\eta_\gamma=0.14\), following the
GeV-afterglow estimate of \citet{Beniamini2015}.

The energy assigned to the full 1~GeV--1~PeV template is calibrated through
\begin{equation}
 E_{\rm HE,true}=\epsilon_{\rm HE}E_{K,\rm true}.
 \label{eq:he_eff}
\end{equation}
The GRB~221009A template gives the isotropic-equivalent energy
\(E_{\rm HE,iso}=3.518\times10^{53}\)~erg.  Applying the
\(0.8^\circ\) TeV-core angle inferred from the LHAASO light curve
\citep{LHAASO2023} gives \(E_{\rm HE,true}=3.43\times10^{49}\)~erg.  The
central narrow-component kinetic energy
\(E_{K,221009A}=5\times10^{50}\)~erg \citep{Sato2025} implies
\(\epsilon_{\rm HE}=0.0686\); the beaming-corrected
\(E_{K,221009A}\simeq4\times10^{50}\)~erg from \citet{Laskar2023} implies
\(\epsilon_{\rm HE}=0.0857\).  The same spectral shape is scaled linearly with
the adopted \(E_{K,221009A}\) in all four scenarios.

For the central calibration,
\(\langle E_{\rm HE,true}\rangle=8.03\times10^{49}\)~erg and
\begin{equation}
 \begin{aligned}
 \dot{\mathcal E}_{\rm HE,0}
 &=\rho_{0,\rm P}\langle E_{\rm HE,true}\rangle \\
 &=5.10\times10^{50}\,
   \mathrm{erg\,Gpc^{-3}\,yr^{-1}}.
 \end{aligned}
 \label{eq:localemissivity}
\end{equation}
The scenarios in Table~\ref{tab:psyche} span
\(\dot{\mathcal E}_{\rm HE,0}=(5.10\text{--}13.7)\times10^{50}\)
erg~Gpc\(^{-3}\)~yr\(^{-1}\).  The population integration changes the
redshift and energy weights, not the template shape or the propagation
responses.

For comparison, the GS luminosity-function average gives
\(\langle E_{\rm HE,true}\rangle=3.92\times10^{48}\)~erg at \(z=0\).
Together with its local rate this corresponds to
\(\dot{\mathcal E}_{\rm HE,0}=3.10\times10^{50}\)
erg~Gpc\(^{-3}\)~yr\(^{-1}\).  The ratio of the local emissivities is 1.64;
the remaining PSYCHE--GS difference comes from their redshift-dependent
weights, as discussed in Section~\ref{sec:psyche}.
\begin{table*}[t]
\caption{Secondary PSYCHE afterglow-energy comparison.  The alternatives are
astrophysical calibration scenarios, not confidence limits.  All peaks occur
in the 18--26~GeV Fermi-LAT bin, and none enters the GS EBL band.}
\label{tab:psyche}
\begin{ruledtabular}
\begin{tabular}{lccccc}
Calibration & $\eta_\gamma$ & $E_{K,221009A}$ (erg) & $\dot{\mathcal E}_{\rm HE,0}^{a}$ & peak cascade/IGRB & peak total/IGRB \\
\hline
Central & 0.26 & $5\times10^{50}$ & 5.10 & 0.066\% & 0.42\% \\
High population $E_K$ & 0.14 & $5\times10^{50}$ & 11.0 & 0.14\% & 0.91\% \\
Higher template efficiency & 0.26 & $4\times10^{50}$ & 6.37 & 0.082\% & 0.53\% \\
Combined high-energy case & 0.14 & $4\times10^{50}$ & 13.7 & 0.18\% & 1.14\% \\
\end{tabular}
\end{ruledtabular}
\begin{flushleft}
\footnotesize $^{a}$In units of $10^{50}$~erg~Gpc$^{-3}$~yr$^{-1}$.
\end{flushleft}
\end{table*}

\subsection{Huang--Banerjee reproduction diagnostic}
\label{app:hb}

The VHE-detectability calculation of \citet{Huang2026} uses the source-count
construction of \citet{Banerjee2021}.  Its luminosity function is defined
through detected GRBs and includes viewing angle in the apparent luminosity
distribution.  It cannot be converted into an intrinsic all-orientation
population by multiplying a detected rate by a single \(1/f_b\) while
retaining apparent \(L_{\rm iso}\); that would mix rate and energy
conventions.  It is therefore retained only as a historical reproduction
diagnostic.

The diagnostic is sensitive to duration and energy-proxy conventions.  Using
the fourth Fermi-GBM-catalog distribution of the observer-frame duration
\(T_{90,\rm obs}\), defined as the interval containing 90\% of the prompt counts,
\(\mu_{\log_{10}T_{90,\rm obs}}=1.476\) and
\(\sigma_{\log_{10}T_{90,\rm obs}}=0.189\) \citep{vonKienlin2020}, lowers the mean
injected normalization by a factor 0.641 relative to the broader distribution
used in the literal reproduction.  Here \(\mu\) and \(\sigma\) are the mean
and standard deviation of \(\log_{10}(T_{90,\rm obs}/{\rm s})\).  The proxy
\(T_{90,\rm obs}(1+z)L_{\rm pk}\), where \(L_{\rm pk}\) is the isotropic-equivalent
peak luminosity, also differs from
\(T_{90,\rm obs}L_{\rm pk}/(1+z)\) when \(T_{90}\) is an observer-frame
duration.  Applying both convention changes moves the diagnostic maximum from
0.7720\% at 18--26~GeV to 0.2367\% at 410--580~GeV.  The model addressed GRBs
observed at Earth; since it cannot be unambiguously generalized to the total
GRB emission in the Universe, it is not used to estimate the population
uncertainty in this work.

\section{Details on the transport models and calculations}
\label{app:technical}
This Appendix specifies the adopted response library and the independent
validation needed to reproduce the reported direct and cascade components.
Version-specific code limitations are stated only to delimit which
implementations were used.
\subsection{CRbeam}
\label{app:crbeam}
CRbeam \citep{Kalashev2023} is used for the adopted propagation result.  No
physics correction to its event-transport algorithm was required in our
tests.  We used a corrected \texttt{--print-tau} diagnostic in which photon energy redshifts continuously during the optical-depth integral.  The stock
diagnostic updated time and redshift while holding the photon energy at its
source value.  This issue is confined to the optical-depth diagnostic; a
propagated control node is unchanged.  The deterministic direct component
uses the corrected optical depth.

The adopted response library consists of three statistically independent
arrays on the same grid of 24 source redshifts, 26 source energies, and 33
observed energies, generated with 300 primary photons per source node and
independent random seeds.  The
direct component is evaluated from the same optical-depth tables rather than
estimated from sparse surviving Monte Carlo events.  Reintegrating the
library with the reference GS weights reproduces the reference GS direct
spectrum to better than \(10^{-4}\) in relative normalization.  After
integration over the full population, the relative standard error of the mean is 0.22\%
for the total intensity and 1.42\% for the cascade in the maximum bin.

In addition, the high-energy extension uses 1000 primary photons per node on a grid concentrated at low redshift and high source energy.  Where it overlaps the main response, the difference is -2.8\% to +5.4\%, with a median absolute difference of about 2.8\%.  These checks establish numerical stability; they are not treated as an astrophysical uncertainty.

\begin{figure}[t]
\centering
\includegraphics[width=\columnwidth,
  trim=11 0 0 0,
  clip]{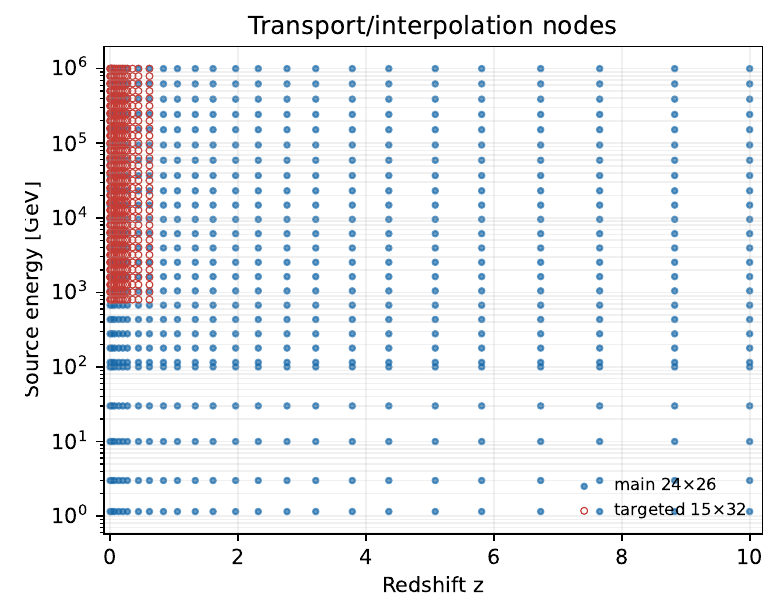}
\caption{Source redshift--energy support of the CRbeam response library.  The
blue points show the main grid of 24 redshifts and 26 source energies; the red
points show the targeted high-energy extension with 15 redshifts and 32 source
energies.  Nodes at \(z>6\) are used only in sensitivity and validation
calculations; the baseline cosmological fold is restricted to \(z\leq6\).}
\label{fig:transportgrid}
\end{figure}

Figure~\ref{fig:transportgrid} illustrates the numerical support used for the interpolations. 

\subsection{CRPropa validation}
\label{app:crpropa}
CRPropa 3 \citep{AlvesBatista2022,Heiter2018} was used only for monoenergetic
tests and auxiliary validation, not for the final cosmological integration.
In the stock 3.3.1 version tested here, the
\texttt{secondariesFirst=True} traversal could skip newly appended secondary
particles when a living parent produced secondaries in several successive
steps.  In an unthinned 100-TeV electromagnetic test this left 20377 active
terminal photons unprocessed.  Using \texttt{secondariesFirst=False} removed
this failure mode.

In the same tested version the inverse-Compton interaction rate was redshift-scaled in \texttt{getRate()} and then multiplied by the redshift scaling once more in \texttt{process()}.  Removing the second multiplication was supported by comparison with CRPropa 3.2.2 and by the  common test with a 100-TeV line injection spectrum.  We regard this statement as version-specific: it refers to the public 3.3.1 source used in this work, not to all CRPropa releases.

Several documented numerical choices also matter for this application.  A
maximum propagation step of 1~Gpc was too coarse for weak absorption at high
redshift; 10~Mpc was required by our survival-probability tests.  An electron
cutoff of 300~GeV suppressed part of the 0.1--1~GeV cascade, whereas 100 and
30~GeV agreed within Monte Carlo precision.  More importantly, the
photon-field redshift treatment used by these modules applies a global scaling
to a reference spectral shape rather than the full two-dimensional evolution
of the Salda\~na--Lopez EBL.  This documented approximation is adequate at low
redshift but became appreciable after integration over the high-redshift
population; hence CRPropa was not used for the final cosmological spectrum.

\subsection{\texorpdfstring{$\gamma$-Cascade}{gamma-Cascade} point-kernel validation}
\label{app:gcascade}
$\gamma$-Cascade V4 \citep{CapanemaBlanco2025} provided the independent
propagation calculation.  For an input spectrum with a hard cutoff, exact
zeros produce \texttt{DirectedInfinity[-1]} under \texttt{Log10} in
Mathematica.  The V4 source handled \texttt{Indeterminate} but not this
exact-zero value at three logarithmic-interpolation steps.  We therefore
locally mapped both cases to a finite logarithmic floor and restored the
corresponding values to zero after interpolation.

The adopted validation avoids the population-level \texttt{*Evolving}
interpolation.  It uses 25 independent \texttt{RedshiftPoint},
\texttt{AttenuatePoint}, and \texttt{CascadePoint} kernels and performs the
redshift integration externally, linearly in flux.  No normalization was
fitted to CRbeam.  The detailed inter-code comparison quoted below uses the
matched \(z_{\max}=10\) sensitivity fold; the sub-percent high-redshift effect
does not affect its interpretation.  This point-kernel calculation gives, for GS, a peak
cascade/IGRB ratio of 0.0178\%, compared with 0.0163\% for CRbeam, and a peak
total/IGRB ratio of 0.1217\%, compared with 0.1069\% for CRbeam.  In matched
integral conventions its total intensity is 6.08\% higher over (0.1--820)~GeV
and 2.58\% higher in (1--10)~TeV.  The cascade-only term differs more: it is
26.78\% higher over (0.1--820)~GeV, with the largest local difference
concentrated near (0.4--1.1)~GeV, while the 1--10~TeV cascade integral is 7.60\%
lower.  Both total spectra remain smooth and monotonic in the TeV range.  We
retain the difference between the two codes as a propagation systematic.

Agreement of the total spectrum alone is not sufficient to validate a
separated cascade component because direct and cascade differences can
partially cancel.  The point-kernel result is therefore compared at the
component level, and the two propagation calculations are not averaged.  For
the GS-only EBL comparison, the Salda\~na--Lopez low- and high-density
calculations use 294 adaptively selected point kernels.  In the four broad
comparison bands, (0.1--1), (1--10), (10--100), and (100--820)~GeV, these kernels
cover (95.00--95.28)\% of the estimated EBL response, and the CRbeam Monte Carlo
standard error is at most 20\% of the resulting EBL-band width.  The Finke
model-A calculation uses 317 kernels.  Twenty-three common nonzero-redshift
kernels were also compared directly between CRbeam and $\gamma$-Cascade.  For
Finke versus the Salda\~na--Lopez central model, the maximum absolute
inter-code morphology difference is 0.00352 for the total and 0.02382 for the
cascade.  For the Salda\~na--Lopez density variants, the cascade passes a 0.10
component-level criterion in all four bands and the total passes 0.02 in
three; the (0.1--1)~GeV total difference is 0.03486.  The maximum missing-kernel
estimate, $\approx 0.003$, and its ratio to the Monte Carlo standard error, $\approx 0.0065$, are much smaller.  This residual is retained as an
independent inter-code morphology uncertainty.

\end{appendix}

\section*{Acknowledgments}
The author thanks O.~Kalashev for early discussions.  This work used the public CRbeam, $\gamma$-Cascade, and CRPropa transport codes. Publicly available large language models assisted the author with language editing and scripting. The author reviewed their suggestions and takes full responsibility for the content of the publication.

\section*{Funding}
This work was supported by the Russian Science Foundation, grant 22-12-00253-P.

\section*{Declaration of competing interest}
The author declares no conflicts of interest.

\bibliographystyle{apsrev4-1}
\bibliography{grb_igrb}

\end{document}